# Growth of Large Area WSe$_2$ and Observation of Photogenerated Inversion Layer in DMOS Configuration

Kajal Sharma[1, #], Abir Mukherjee[1, #], Kritika Bhattacharya[1], Biswarup Satpati[2], Dhiman Mallick[3] and Samaresh Das[1,3]*

[1] Centre for Applied Research in Electronics, Indian Institute of Technology Delhi, 110016, India

[2] Saha Institute of Nuclear Physics, A CI of Homi Bhabha National Institute, 1/AF, Bidhannagar, Kolkata 700064, India

[3] Department of Electrical Engineering, Indian Institute of Technology Delhi, Delhi, 110016, India

* Corresponding author: samareshdas@care.iitd.ac.in

#: KS and AM contributed equally in this work

## Abstract

Here, we report a full-fledged journey towards the material synthesis and characterization of a few-layered/thin WSe$_2$ using sputtered W-films on SiO$_2$/Si substrates followed by electrical studies under dark and illumination conditions. Growth temperature of 500°C and a gas flow rate of 55 sccm are found to be the optimized parameters for the formation of thermodynamically stable WSe$_2$ with dominant Raman peak at 265 cm$^{-1}$. XRD and HR-TEM measurements clarify the formation of high crystallinity along the c-axis and quasi-crystallinity along a and b axes, respectively. Lower intensities from Raman-measurement and PL-peak at 768 nm (with 532 nm excitation wavelength) infers the thin nature of the grown film. This work also re-tracks the controlled etching by reactive ions to achieve large area bi/tri-layer films to fabricate high performance devices.  An advanced dual MOS (DMOS) structure on SiO$_2$/p-Si substrate is fabricated which shows tremendous performance by means of photo-capacitance under illumination condition where photo-carriers can survive the higher probe frequencies. Under illumination conditions, the DMOS device demonstrates superior performance, exhibiting a significantly strong electron-inversion region compared to HfO$_2$/SiO$_2$/p-Si and SiO$_2$/p-Si MOS devices, even at high frequencies (1-10 MHz).  This work thus presents a potential approach for capacitance-based, highly sensitive photodetection within conventional Si technology, enabled by integrating WSe$_2$/W as the active material.

**Keywords:** Transition Metal Dichalcogenides, Chemical Vapor Deposition, Reactive Ion Etching, Density of States, MOS device

# 1. Introduction

Over the past few decades, technology based on electronic devices has brought revolution in all possible fields, starting from society to space research. Still, Si, Ge, GaAs, etc. materials are the key semiconductors for electronic and optoelectronic devices. Therefore, these are well explored from a material point of view and challenging as well for new and fundamental research. Atomically thin 2D materials open doors to countless applications across diverse fields, from electronics to medicine, due to their exceptional properties. First, quantum confinement in the direction perpendicular to the 2D plane leads to unique electronic and optical properties which make them different from their bulk counterparts. Second, the high surface-to-volume ratio ensures that many atoms, possibly half or more, are at or near the surface, making them different from bulk systems. In the past few years, 2D materials such as graphene, TMDCs have grabbed significant attention in the nano/quantum electronics field due to multiple easily tunable properties, e.g., mechanical, optical, and electrical[1–5]. Transition metal dichalcogenides (TMDCs), characterized by the formula $MX_2$, where M represents group IV–X transition metals (Mo, W, Ti, V, Nb, Hf etc.) and X denotes chalcogens (S, Se, and Te), thereby introducing a large family of materials with tunable electronic and topological properties[6–8]. For example, band gaps of semiconducting TMDCs can be tuned by controlling layer numbers, growth processes, chemical compositions, strains of the materials, etc. On semiconductor TMDCs, significant efforts are being conducted toward synthesis, structure, and defect characterization[9,10]. Among the various semiconducting TMDCs, $MoS_2$ initially garnered the most attention. Over recent years, there have been numerous reports on the growth of large-area crystals, continuous films, and patterned monolayer and few-layer $MoS_2$ structures. Moreover, field effect transistors (FETs) fabricated using both mechanically exfoliated and vapor phase-grown $MoS_2$ have been widely studied[11,12]. Generally, $MoS_2$ has an n-type nature, although p-type characteristics have also been reported in special cases. Later, $WSe_2$ was introduced as a unique complementary and a better choice from an electronic to optical point of view due to its relatively high mobility, spin-valley selection, lowest thermal conductivity, etc. Monolayer $WSe_2$ has a band gap smaller than monolayer $MoS_2$ (~1.65 eV for monolayer $WSe_2$ versus 1.8 eV for monolayer $MoS_2$), while they possess similar band gaps in bulk form (1.2 eV for both)[13–15]. Importantly, it has already been demonstrated that the transport properties of mechanically exfoliated monolayer $WSe_2$ can be easily tuned to be either p-type or ambipolar behavior, depending on the types of contact metals[16–18]. $WSe_2$ has a huge absorption coefficient in the visible to infrared range, a high quantum yield in photoluminescence (PL), and a strong spin-orbit coupling (heavy SOC in bulk)[19]. Leveraging split gate device structures and the above advantages, a variety of optoelectronic devices such as photodetectors, light-emitting diodes, and photovoltaic devices have been demonstrated using mechanically exfoliated monolayer $WSe_2$. Among other TMDCs (semiconductor & semi-metal) such as $MoS_2$, $WS_2$, $MoTe_2$, etc., the growth of $WSe_2$ holds huge challenges and deserves finely controlled growth mechanisms in Chemical Vapor Deposition (CVD) and Molecular Beam Epitaxy (MBE) since selenium precursors are less reactive than sulfur precursors. In few recent studies, a huge focus has been provided to eliminate the growth regarding challenges for $WSe_2$ [20–23]. In the cases of $MoS_2$ and $WSe_2$, $WO_3$ is much more difficult to sublimate than $MoO_3$ due to their large difference in boiling points and, consequently, vapor pressures (the boiling points of $WO_3$ and $MoO_3$ are 1700 ºC and 1155 ºC, respectively). As per different articles, vapor phase-grown $WSe_2$ usually exhibits either p-type or ambipolar transport behavior, while tunability of such transport behavior in the same material has rarely been reported. CVD is the best technique for bottom-up approaches for synthesizing the material, and for that, we can use two different approaches; in the first approach, we can pre-deposit the transition metal source on the growth substrate using sputtering/e beam evaporation followed by the selenization of the film under requisite gas pressure[24]. The second approach is to use both the metal and selenium source in precursor form with gas flow from one end to synthesize the TMDC material on the growth substrate[25,26]. The second approach is obligatory for the growth of ultra-fine structures such as mono/bi/tri-layer, but the first approach is much more promising for large-area growth which has been broadly discussed in this work. Here, we report a large area growth of semiconducting $WSe_2$ on the $SiO_2$/Si substrate and, thereafter, the Reactive Ion Etching (RIE) of the film to a few layers to make it applicable for advanced devices (e.g. photodetector, FET/HEMT, LED, Quantum Emitter etc.). In this manuscript, this grown film is implemented to form an advanced Metal-Oxide-Semiconductor (Dual MOS, abbreviated as DMOS) structure, which shows tremendous performance in terms of the photogenerated electron inversion layer, even at high frequencies (1-10 MHz). Such a device is useful in various applications, particularly in areas that benefit from enhanced photodetection, charge separation, and fast response to light.

## 2. Results and Discussion

**Fig. 1(a)** shows the schematic of the chemical vapor deposition (CVD) furnace setup, in which the $SiO_2$/p-Si substrate with a tungsten (W) film (27±2 nm) was placed for the selenization process (details of the synthesis process are provided in the Methods section). Initially, the deposition of the W film via RF sputtering was optimized. Since RF power plays a very crucial role in the film's uniformity, it has been varied in this deposition process for proper plasma sheath formation. It has been observed that deposition with lower power (< 60 W) causes non-uniform film, therefore being inappropriate for the CVD-synthesis process. After proper analysis, an optimized RF power of 100 W was chosen for the deposition of W film. Selenium being extremely less reactive, it is challenging to find reaction temperature with W-film. The Se-powder melting temperature starts from 280°C, but it can vary up to 350 °C for better optimization under different process conditions (atmosphere/vacuum/plasma). **Fig. 1(b)** shows the schematic representation of the grown film over the deposited W, depicting how the film grows during the selenization process. Initially, W-film is pre-oxidized forming $WO_{3-x}$, which then acts as a precursor for forming $WSe_{2-x}$. A combination of argon (Ar) and hydrogen ($H_2$) gas is utilized in the growth process, with Ar serving as an inert atmosphere and carrier gas, while $H_2$ aids in removing oxygen from the film to facilitate the formation of $WSe_x$. The transition from $WSe_x$ to $WSe_2$ is possible after proper post-annealing mechanism. Post-annealing or high-temperature growth (400°C <$T_{Growth}$<550°C) allows the film to achieve better crystallinity along the c-axis, as shown in Fig. 1(b). It is preferable to avoid increasing the growth temperature beyond 550°C, as this may lead to excessive diffusion of Se atoms into the underlying W-film, thereby compromising the crystalline quality (structural order). Moreover, the growth temperature (GT) should not exceed 600°C, as higher temperatures can cause evaporation and removal of the deposited thin W-film due to its weak binding with $SiO_2$-layer outside the CVD chamber. Conversely, the GT should not be set below 400°C, as it would be insufficient to oxidize the W-film, which acts as a solid-phase precursor in this process. Therefore, in this work, GT is tuned from 400-550°C. **Fig. 1(c-f)** shows the optical images of the grown film at different GTs viz. 400°C, 450°C, 470°C, and 500°C with inset showing the corresponding Electron Probe Microanalysis (EPMA) mapping. According to the EPMA analysis, the average mass percentage for W is found to be 10 times higher than Se which refers to the fact that only 10% of deposited W is being selenized with a thickness of ~12±2 nm, and rest of beneath-W remains un-selenized.

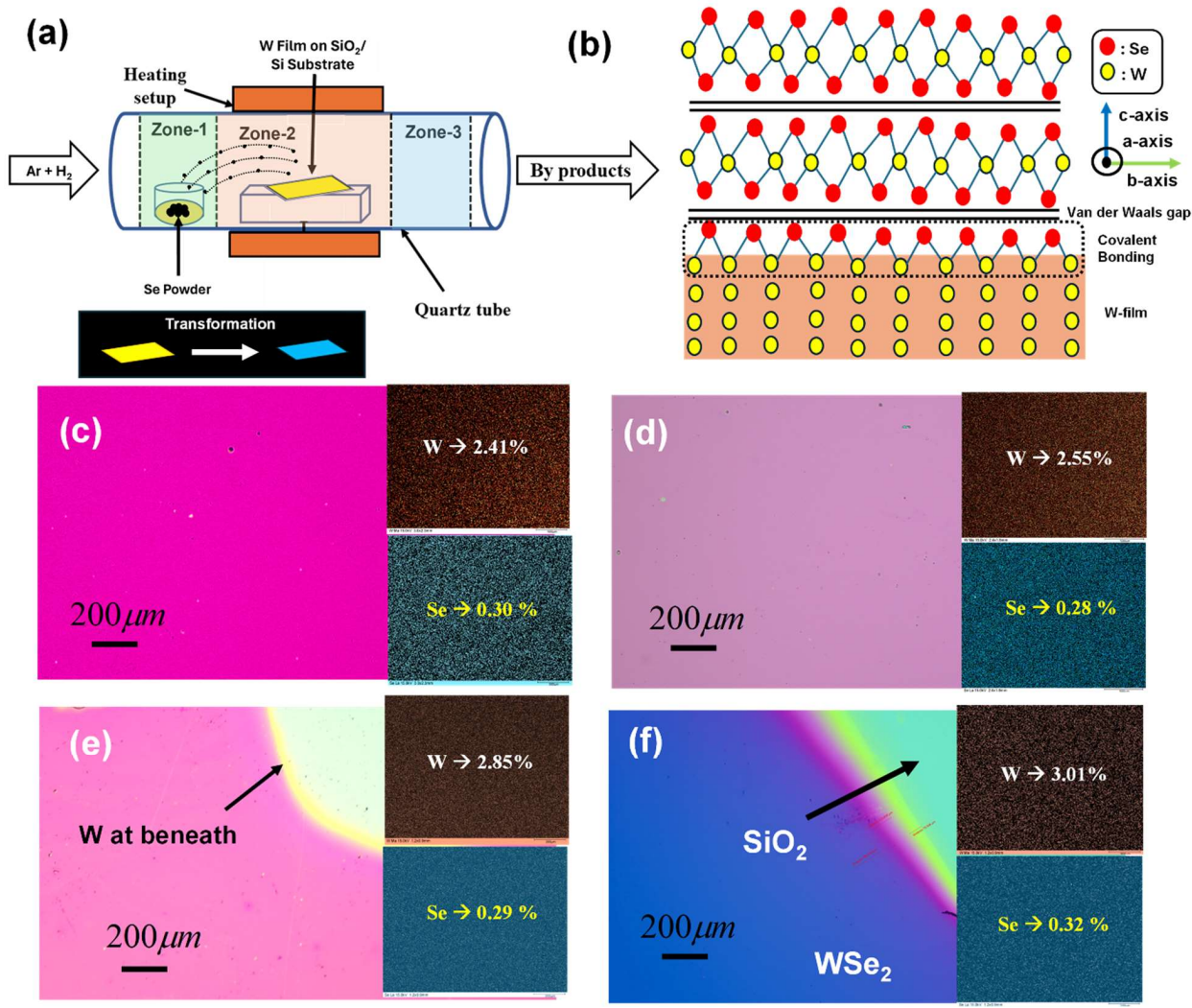

**Fig.1.** Schematic representation of the (a) CVD set-up, (b) formed WSe$_2$ film over the RF sputtered W; Optical images of the WSe$_2$ film at different growth temperatures viz. (c) 400 ºC, (d) 450 ºC, (e) 470 ºC, (f) 500 º with corresponding EPMA mapping for W and Se.

Raman spectra, as shown in **Fig. 2(a)**, depict the Raman peaks of the WSe$_2$ film grown at different temperatures. A dominant peak is observed at 265 cm$^{-1}$ for all the grown samples. Although the Raman shift may vary for different bond lengths and orientations, the dominant peaks of WSe$_2$ may vary from 240-270 cm$^{-1}$ [24,27–31] where two dominant peaks, **A$_{1g}$** and **E$^1_{2g}$** correspond to out-of-plane vibrational and in-plane vibrational modes, respectively. For bulk WSe$_2$, regardless of whether it is synthesized via CVD or obtained through the mechanical-exfoliation, either a single peak or two distinct Raman signals are observed, consistent with literature[32]. For a few layered WSe$_2$, as in this work, a single maxima is found, the position of which changes with the number of layers. Currently, it is unclear whether the single maxima is due to which mode because the two modes are nearly degenerate and cannot be resolved. Notably, a broadside maximum at 260 cm$^{-1}$ is found for bi-layer[30,31]. However, it can be concluded that the presence of Raman peaks except at 250 cm$^{-1}$ and nearby may originate due to the formation of phase/bond/layer-modulated WSe$_2$ with varying electronic and optical properties. Moreover, it is noteworthy to mention that such weaker Raman-intensities, even after 3-4 accumulations, indicate the formation of thinner WSe$_2$ and shift towards higher frequencies (e.g., 260→268 cm$^{-1}$) depicts the reduced bond length[33,34] between W and Se and formation of multi-layer[35] as shown in **Fig. 2(a)**. Such blue shifts of the peak compared to the pristine WSe$_2$ indicates p-doping in WSe$_2$ and a well-resolved peak nearly at 310 cm$^{-1}$ (as depicted in the Raman spectra in **Fig. 2(a)**) arises due to W-O-W stretching vibration, indicating the presence of WO$_{3-x}$ phase[36]. It is clearly distinguishable from **Fig. 2(a)** that WO$_3$ is the dominant material over WSe$_2$ at lower temperatures (@ 400 °C), whereas WSe$_2$ achieves its dominance at higher ones (e.g., 500 °C). Therefore, XRD and Raman data can be comparatively tallied for better insight into material chemistry. Post-annealing or high-temperature growth (400°C <T$_{Growth}$<550°C) allows the film to achieve better crystallinity along the c-axis, which can be visualized from the XRD plot whereas the crystallinity along the a and b-axes is observed via RHEED-pattern (with 1 mm spot size) and it is found to be quasi/poly-crystalline. Formed WSe$_2$ exhibits high crystalline properties along the c-axis, where peaks are compared with the XRD peaks of W-film[37,38]. The PL spectra reveal a strong direct transition emission around 760 nm (1.631 eV) in the monolayer (ML) due to its direct band gap. With an increasing number of layers, additional indirect transitions emerge at lower energies, which are absent in ML PL spectra. For WSe$_2$ grown at various temperatures, in this work, the PL data show a dominant peak at 768 nm (1.614 eV), indicating a redshift from 760 to 768 nm (**Fig. 2 (c)**) with 532 nm excitation wavelength. This shift suggests an indirect transition characteristic of the few-layer (~10 nm) structure, where the magnitude of this shift can serve as a layer number indicator. In the high vacuum low-temperature PL experiment conducted for the film grown at 500°C, significant diminution in the intensity of emission spectra has been observed with a blue shift towards 760 nm, which infers the dominance of direct-transitions including the excitonic impacts[30,37] at ultra-low temperatures (e.g., 10 K) as shown in **Fig. 2(d)**. Alongside, the film exhibits strong second harmonic emissions observed with excitation wavelengths 350 nm to 450 nm. X-ray photoelectron spectroscopy (XPS) was conducted to acquire the binding energies of W 4f and Se 3d. For W 4f core level spectra (**Fig. 2(f)**) at optimized GT: 500°C, W $4f_{7/2}$ and W $4f_{5/2}$ peaks appear at 33.62 eV and 35.72 eV, respectively, while Se $3d_{5/2}$ and Se $3d_{3/2}$ peaks are at 53.01 eV and 54.73 eV, respectively, corresponding to W-Se bond (rescaled after carbon correction). XPS measurements were found to exactly similar for other growth temperatures. It is noteworthy to mention that left shifting (i.e., towards higher binding energy (B.E.)) is attributed to the decrement in the electron density around the nucleus, whereas lower binding energy corresponds to the enhancement of the same. Such shifts in B.E. may originate from electron-donating effects or a less electronegative environment[39]. Ideally, in XPS-measurement, the Se-$3d_{5/2}$ peak comes at around 54 eV, whereas here, the peak shifts to 53.01 eV, causing a right shift as shown in **Fig. 2(e)**, indicating the existence of natural strain in the material[40]. The W 4f and Se 3d doublets can be categorized into two groups, corresponding to the 2H and 1T phases, which are known to be semiconducting and semi-metallic respectively. [24,41,42]. XPS analysis was done on different spots to verify the uniform growth with varying beam energies, spot area and analysis time. Peak $4p_{3/2}$ again validates the existence of $WO_{3-x}$. From **Fig. 1(b)**, it is possible to visualize that the exact first layer of $WSe_2$ or $WSe_x$ above the W-film is hardly bound with covalent bonding, whereas the upper layers are separated by van der-Waals gap, thereby bi-layer/ monolayer can be transferred via wet/dry transfer method[18,43] to another substrate for high mobility transistor (HMT) and/or phototransistor (PT) applications. Raman map was taken that refers to the fact of less material spatial-anisotropy, i.e., better surface homogeneity along the a and b-axis, to validate the schematic representation of **Fig. 1(b)**.

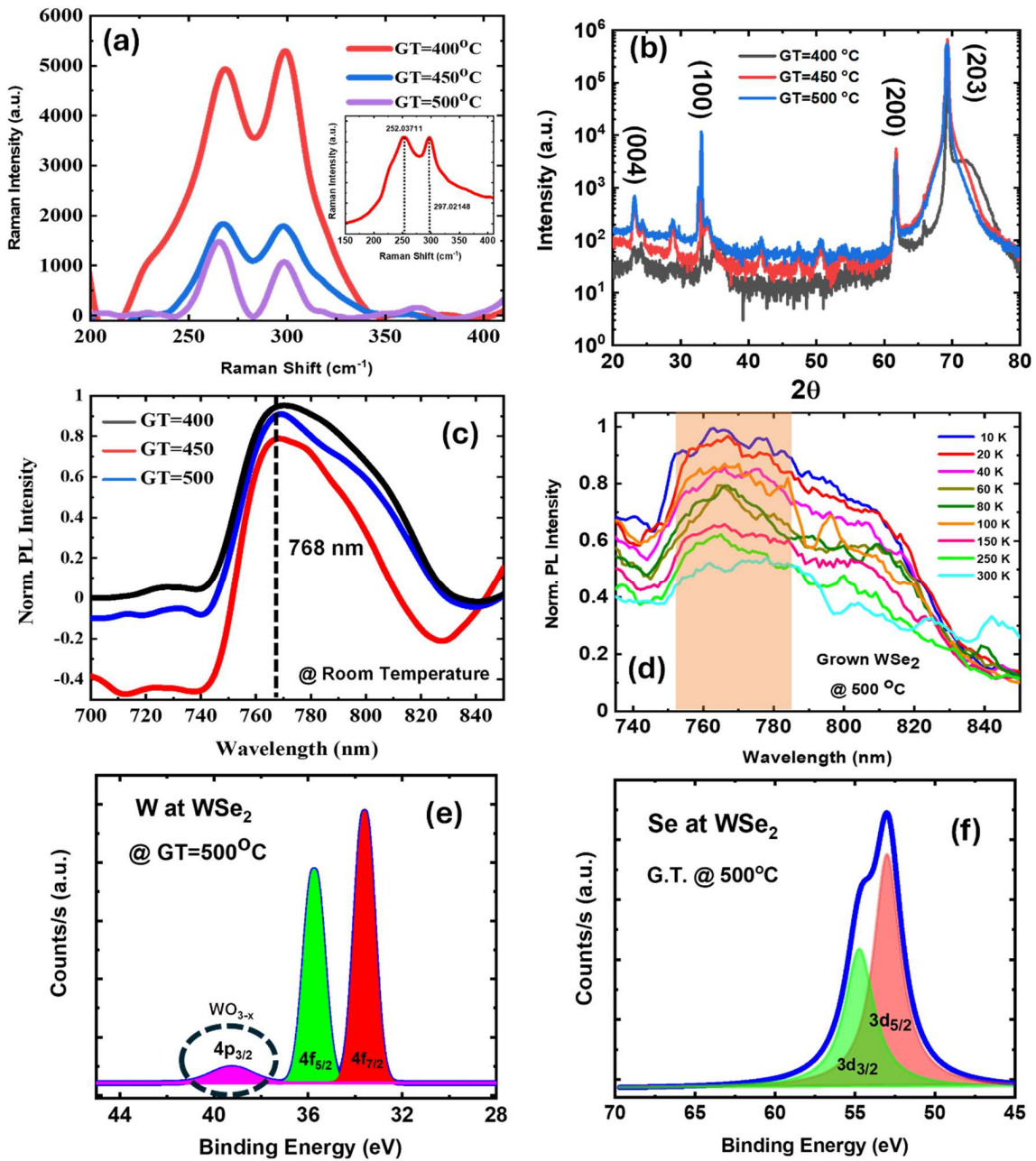

**Fig.2. (a) Raman Spectroscopy of grown WSe$_2$ at different temperatures with three accumulations, (Inset depicts the conventional Raman peak after post annealing at 400°C with growth temperature 500°C), (b) XRD Spectra of the grown films, Normalized (c) room temperature, (d) low temperature PL spectra, (e-f) Normalized XPS spectra with varying growth temperatures.**

The surface morphology of the grown film was studied very accurately via Scanning Tunneling Microscope (STM) within the ultra-nanoscale region, as shown in **Fig. 3(a)**, though the intention was to visualize the atomic orientations. Evidently, all the above results indicate the formation of a few layered/thin $WSe_2$[44]. To confirm the exact thickness of the grown film, RIE was performed after patterning the sample using photolithography and forming a window (details are discussed in the Methods section). The optical and SEM image in **Fig. 3(c-d)**, respectively clearly illustrates the etching profile, displaying areas of the grown film, partially etched film, and fully etched film, exposing the underlying tungsten. **Fig. 3(b)** shows the SIMS-depth profiling (DP) that depicts the W and Se distribution by which the selenized (i.e. formed $WSe_2$ region ~14 nm) and unselenized region (i.e. remaining beneath W) can be resolved. (**see Supporting Information**), as also confirmed by XPS suggesting that the grown film nature is intrinsically p-type[45]. AFM was carried out clarifying the grown film thickness to be around ~14±2 nm, with roughness comparison with W-film as shown in **Fig. 3(e)** along with the scope of thinning down up to a few layers (e.g., bi/tri-layers) by controlled etching using RIE. It is noteworthy to mention that $SF_6$ also etches out W[46] which causes little inaccuracy in the calculation of $WSe_2$ thickness; therefore, we tried to validate the thickness with XRR measurement and FESEM cross-sectional images and found them to be consistent with the AFM results. Alongside, $WO_{3-x}$ was also found beneath instead of W for high temperature growth. KPFM measurement was carried out to extract the surface potential of the film (**Fig. 3(f)**). The surface potential was found to be bound between 488 meV and 422 meV; thereby, the calculated work function of $WSe_2$ varies in the range of 4.652 to 4.718 eV (with tip work function ~ 5.14 eV).

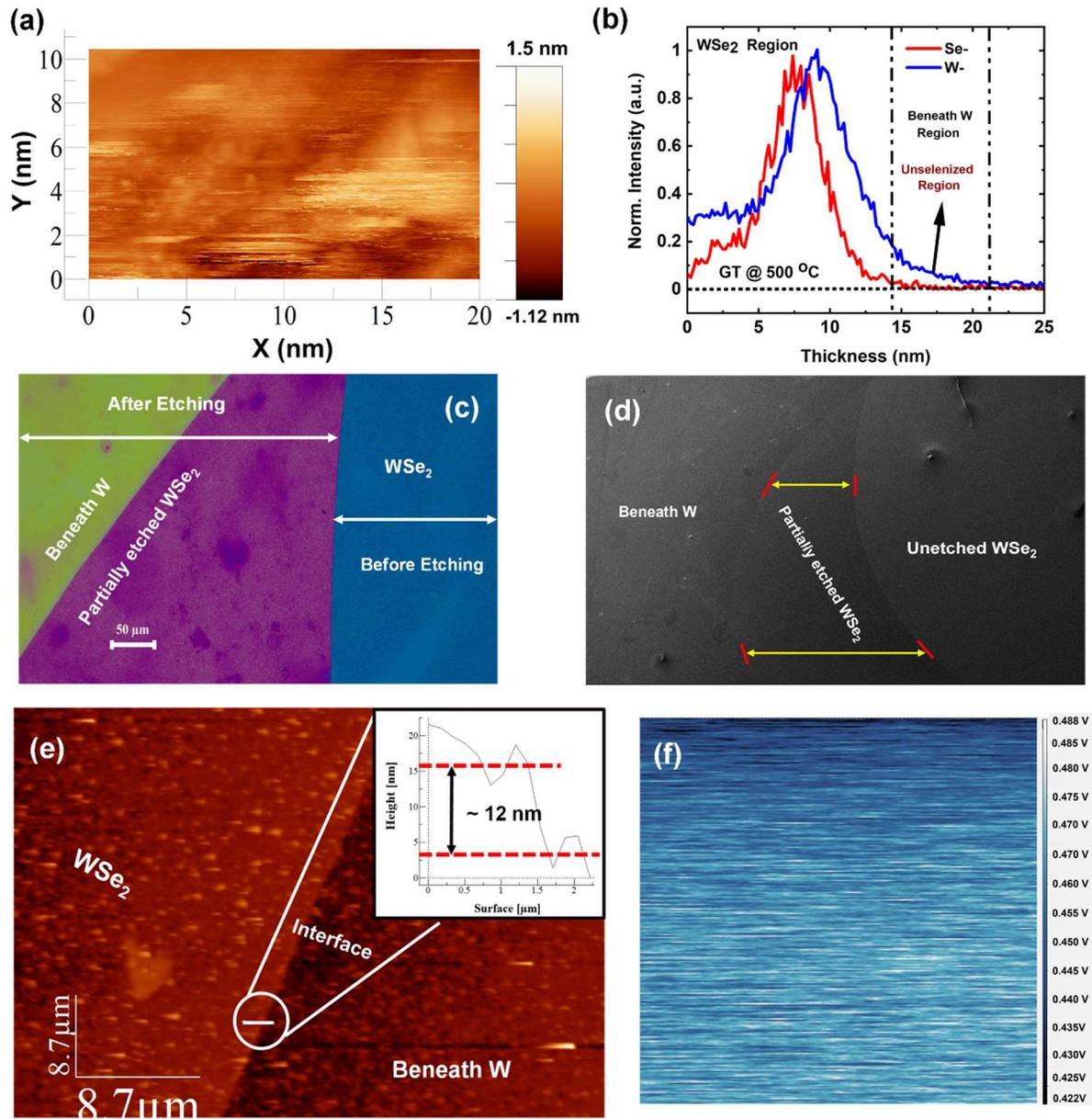

**Fig.3.** (a) STM image depicting the surface morphology of grown WSe$_2$ at ultra nanoscale, (b) TOF-SIMS depth profile, (c) Optical microscope image of reactive-ion etched film, (d) SEM top-view image of the film, (e) AFM image with thickness calculation, (f) KPFM measurement data on the grown film @ 500°C.

To investigate the critical impact of the underlying W layer, particular attention is given to its electronic density (local density) of states. The semiconductor nature of WSe$_2$ arises from fully occupied Se-4p shell which energetically lies below the W-5d states. The remaining electrons completely fill the W- $d_z^2$ orbital in which the trigonal crystal field gets split off from the rest of the d-manifold, thus opening a gap between occupied states (valence band) and unoccupied states (conduction band). The crystal structure of 2H-WSe$_2$ belongs to the D$_{6h}^4$ space group with lattice parameters: a=b=3.325 $\overset{0}{A}$ (in plane) and c=10.46 $\overset{0}{A}$ (perpendicular).

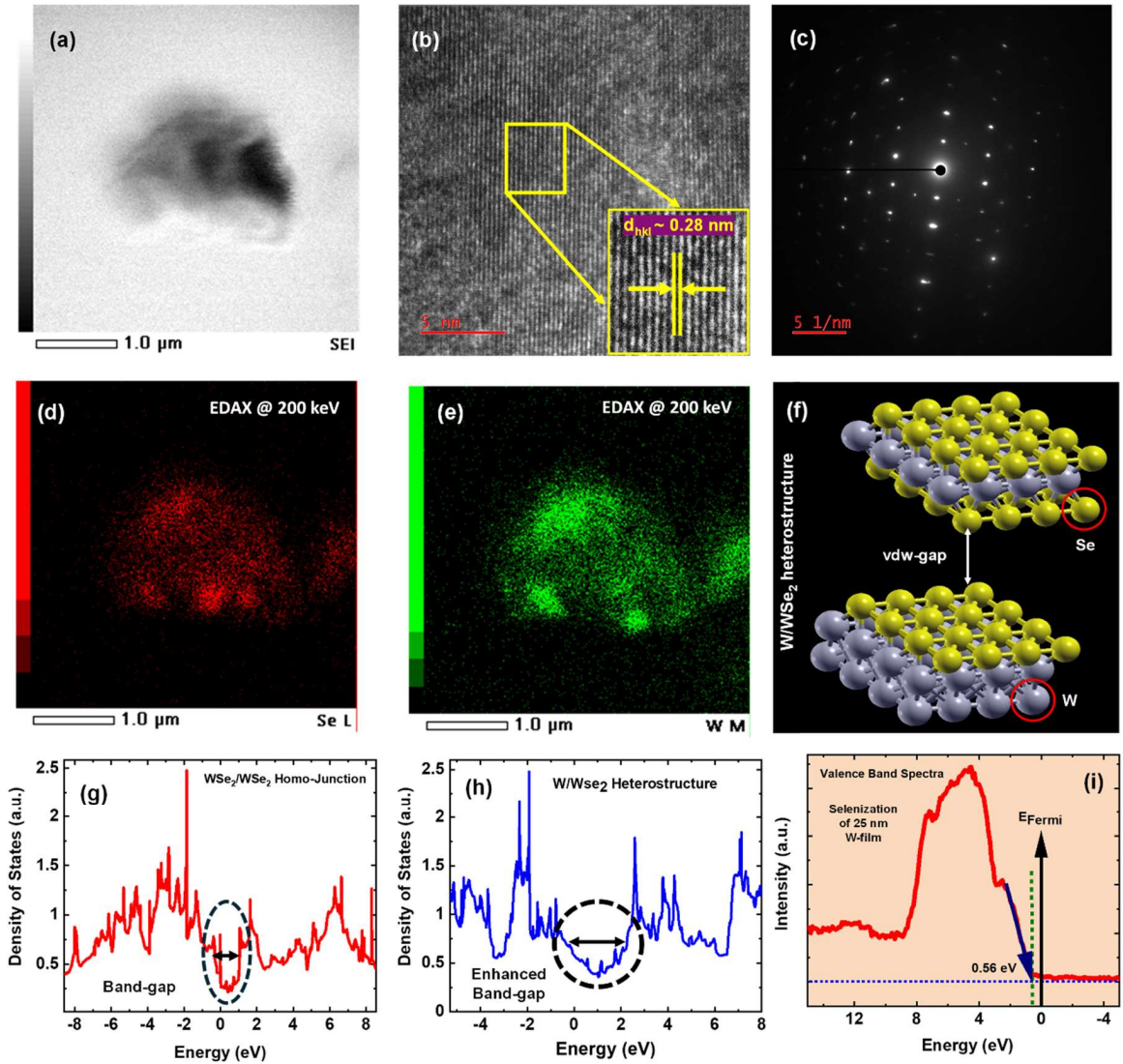

**Fig.4 HR-TEM of the WSe$_2$ film (a) at low, (b) high magnification, (c) corresponding SAED pattern, Elemental mapping set up (d) Se & (e) W using HR-TEM, (f) designed heterostructure as per growth, local density of states (g) of conventional bilayer, (h) W/WSe$_2$ heterostructure to convey the impact of beneath W, (i) valence band spectra using XPS set up**

In this work, crystal structure has been considered in such a way that Se-diffusion reaches down to some extent at W-lattice forming WSe$_x$ which is in vander-waals (vdW) gap from the exact next layer of WSe$_2$. It is noteworthy to mention that the existence of WO$_{3-x}$ has not been incorporated.

From **Fig. 4(h)**, it is clearly visible that the beneath W introduces swallow trap states below the conduction band along with increasing the bandgap to ~ 1.75 eV (compared with respect to bi-layer band gap ~ 1.35 eV as shown in **Fig. 4(g)**). To visualize the crystal lattice, HR-TEM was performed where crystalline plane, (100) was found to follow with d-spacing of 0.28 nm (**Fig. 4(b)**) with corresponding elemental analysis and SAED patterns **(Fig. 4(c))**. SAED patterns for different locations can be found from **Supporting Information**. A detailed analysis of the SAED pattern confirms the formation of highly crystalline planes along the a- and b-axes. EDAX was performed on the sample as shown **Fig.4 (a,b)** (at lower & higher magnification respectively) to confirm the existence of W and Se as represented in **Fig.4 (d, e).** Valence band spectra measurement again confirms the p-type nature of the film, since the edge of the valence band and fermi energy level are gapped by approximately 0.56 eV. It was also found that grown film via selenization of thinner W ($\leq$ 10nm) shows n-type characteristics (**see Supporting Information**).

To characterize the intrinsic doping properties and photo response of these CVD-grown films, capacitance measurements were conducted on fabricated dual-Metal oxide Semiconductor (DMOS) structures (Au/HfO$_2$/WSe$_2$/W/SiO$_2$/p-Si), (details of the fabrication process is discussed in the Methods section) as shown in **Fig.5(a)**. Moreover, MIS or MOS structure provides the scope to scan the local density of states by means of capacitance. The existence of underlying W may serve as a charge accumulation region; therefore, dual sweep measurements were carried out for capacitance, both under dark and illumination conditions to validate the existence of any hysteresis window.

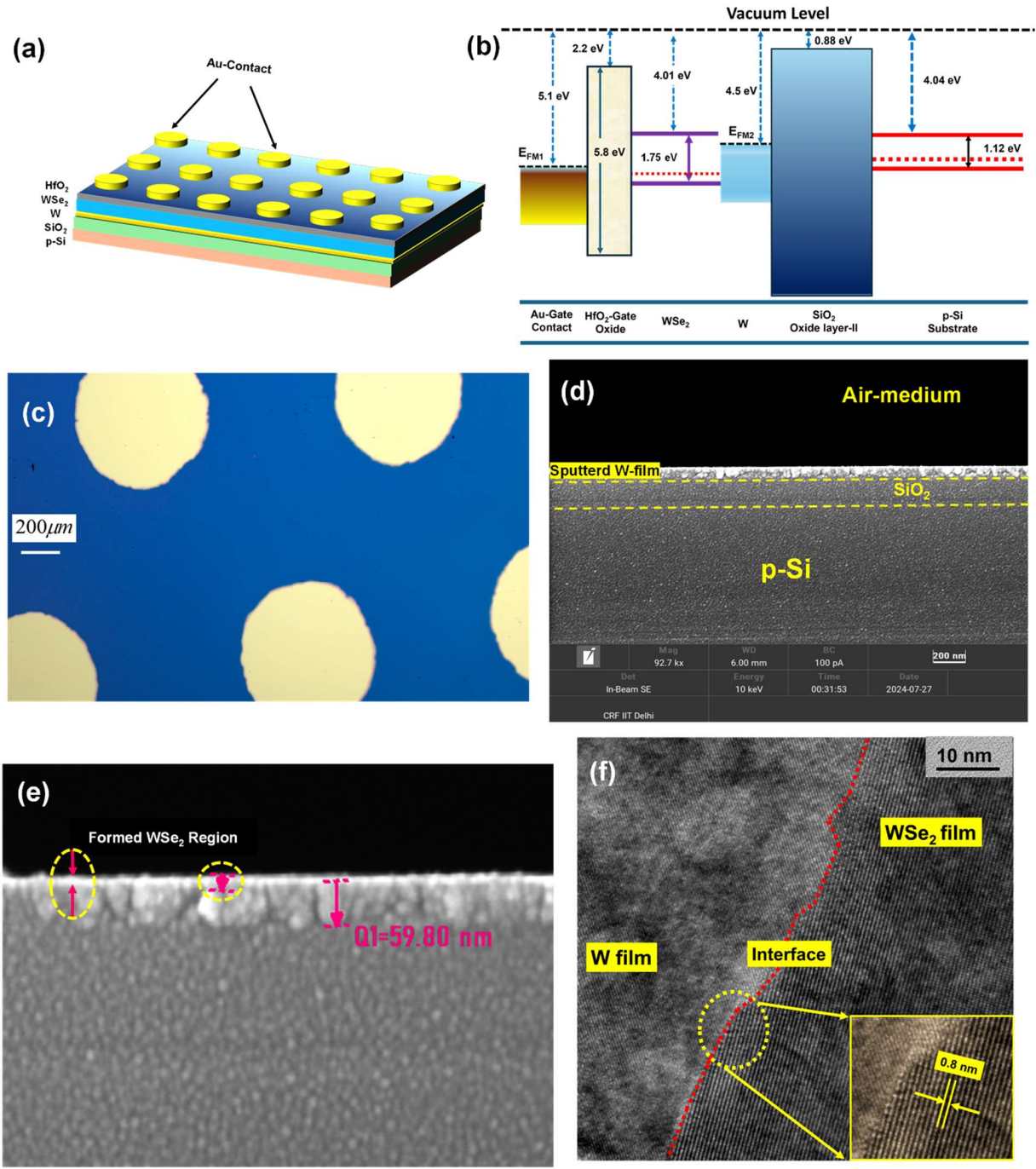

**Fig.5** DMOS device (a) schematic representation, (b) band diagram, (c) Optical microscopic image of Fabricated Patterned Array, (d, e,) cross-sectional FESEM image of the sequentially grown materials with (f) cross-sectional TEM image of WSe$_2$/W heterostructure

**Fig.5(b)** presents the energy band diagram, illustrating the energy levels of each layer within the DMOS structure. **Fig. 5(c)** displays the optical image of the fabricated device. Additionally, the cross-sectional FESEM images distinctly show the individual layers of the structure, as illustrated in Fig. 5(d-e), while Fig. 5(f) presents the interface of $WSe_2$/W heterostructure via high resolution annular bright field scanning cross-sectional transmission electron microscope (STEM) image which shows clean interface vdW interface between $WSe_2$ and W. Such clean interface material based devices are of huge demand for spin injection applications[47].

Since films grown at 450°C and 500°C exhibit nearly identical characteristics with uniform growth and better crystalline properties, therefore these were chosen for C-V measurements under dark and illumination conditions (with laser source 660 nm and 785 nm) for a wide range of probe frequencies (10 kHz to 10 MHz) and gate-bias: -30V to +30V. To initiate the discussion on C-V characteristics, it is essential to understand the underlying device physics. The W-film beneath the $WSe_2$ layer acts as a bulk trap, pinning the Fermi level of $WSe_2$. When a positive gate bias (>10 V) is applied, substantial band bending occurs, leading to deep electron inversion. Although a deep triangular inversion potential well forms within the conduction band, it remains only partially occupied. Under laser illumination, sub-band filling occurs due to photo-doping[48–50]. **Fig.6(a-h)** demonstrates that, under dark conditions, only a weak electron inversion (EI) forms. However, with optical illumination, the capacitance catches a significant enhancement even at higher probe frequencies, indicating that the photogenerated carriers attain extended lifetimes due to the deep inversion potential well. The observation of significant inversion at higher frequencies in the fabricated DMOS device suggests a high-quality $HfO_2$/$WSe_2$ interface. Since such inversion is not achieved for Au/$SiO_2$/p-Si or Au/$HfO_2$/$SiO_2$/p-Si MOS devices under dark and laser illumination conditions, that refers to the fact that the intermediate $WSe_2$/W or $WSe_2$/$WO_3$ layer contributes to achieving huge photo-capacitance (PC) even at higher frequencies. Finally, from **Fig.7 (a,b)**, it is possible to conclude that capacitance under positive bias originates from photo-generated electrons in the deep triangular inversion potential well, which remains completely omitted in the case of other MOS devices. In the C-V curves for samples grown at 500°C, the peak inversion capacitance is approximately equal to the peak accumulation capacitance at a frequency of 50 kHz under illumination, with a slight decrease observed at higher frequencies, as shown in **Fig. 6(a, c, e, and g)**. In contrast, for the sample grown at 450°C, the maximum inversion capacitance is sufficiently lower than the accumulation capacitance (**Fig. 6(b, d, f and h)**).

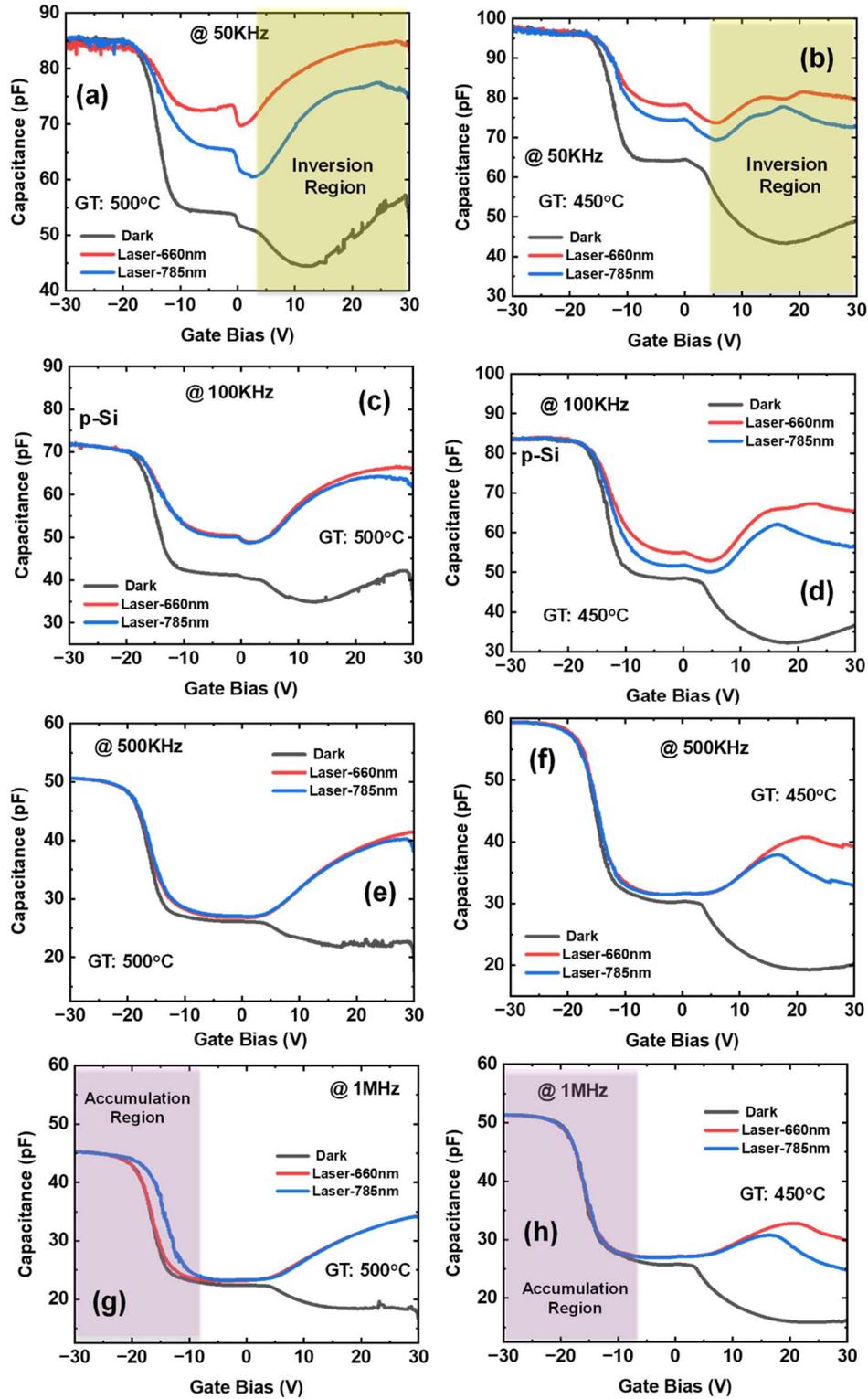

**Fig.6** DMOS Capacitance vs Voltage plots for grown WSe$_2$ (500°C and 450°C) from low to high probe frequencies (a-h), respectively under dark and illumination condition (using 660nm and 785nm wavelength)

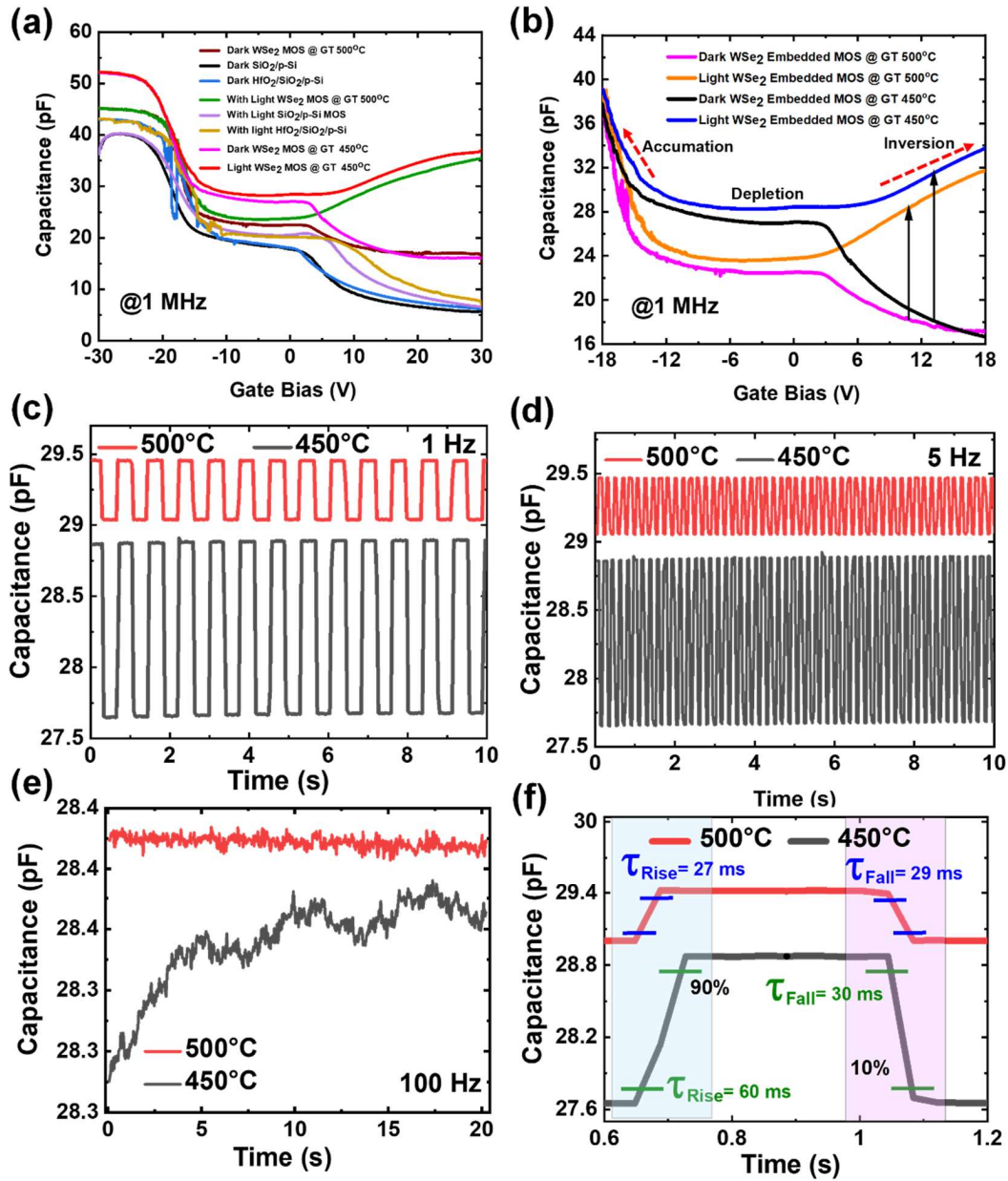

**Fig.7** Capacitance vs voltage plot of (a) all fabricated MOS devices, (b) the fabricated DMOS device grown at 450ºC and 500ºC showing the accumulation, depletion, and inversion region under dark and illumination conditions at probe-frequency of 1MHz. Dynamic photo-response of the fabricated DMOS device at optical modulation frequencies of (c) 1 Hz, (d) 5 Hz, (e) 100 Hz. (f) Rise and fall time calculation.

Furthermore, as discussed, hysteresis was seen in both the samples, which signifies the trap states inside the samples. However, it is noticed that the sample grown at 450°C has more hysteresis area. Therefore, it may be concluded that comparably, it has more interface or bulk (due to beneath W) traps compared to 500°C, which again infers better crystalline properties of the 500°C grown sample. The existence of hysteresis window (HW) can be broadly studied in future work to realize any light-activated memristor characteristics[51,52] since HW broadens under optical illumination dominantly at lower frequencies.

To understand the dynamic response of the fabricated DMOS device, it was subjected to a constant bias of 0.5 V. At this bias, the device was in depletion mode. Under the light modulation of 1 Hz, capacitance displayed a significant rise and fall. Since the MIS or MOS capacitor was in the depletion region, the generated additional photo-carriers did the job of reducing the depletion width, therefore effectively increasing the capacitance. When the light was turned off during modulation, the generated carriers started to recombine, and the capacitance began to return to its original value as shown in **Fig.7(c)**. As can be seen from **Fig.7(d)**, the capacitance weakly followed the trend of increasing the modulation frequency of the optical signal to 5 Hz, which completely diminished at 100 Hz (**Fig.7 (e)**). The rise and fall times were calculated to be 60/27 ms and 30/29 ms for samples grown at 450/500°C, respectively (**Fig.7(f)**). It is very much requisite to have better insight into the dynamic response at 100 Hz, which suggests that photo-carriers have a much higher lifetime than the optical modulation frequency; therefore, PC kept on increasing and got saturated after a certain time, whereas in other cases, PC followed the lesser modulation frequency, infers the equivalent and lesser carrier lifetime.

## 3.     Conclusion

In this work, we have demonstrated the growth of thin large area $WSe_2$ with further possible minimization to ultrafine structures (mono/bi/tri-layer) by reactive ion etching. Moreover, this growth mechanism offers formation of highly photo-sensitive film along with strong second harmonic generation applicable for various quantum, THz and advanced microscopic applications. Features provided by controlled partially etching gives an opportunity to fabricate advanced HMT devices utilizing layer/thickness induced property tuning. This growth mechanism also nullifies the constraints regarding the growth of large area $WSe_2$. An advanced dual MOS (DMOS) structure demonstrated tremendous performance by means of photo-capacitance under illumination conditions even at higher probe frequencies. Formation of metal induced trap level hints towards memory applications with optical switching. Such a device holds the potential for capacitance-based, highly sensitive photodetection within conventional Si technology, enabled by integrating WSe$_2$/W as the active material. Capacitance versus time measurement enables the route to directly estimate carrier lifetime.

## Methods

### 1. Deposition of Tungsten film & CVD-Growth of WSe$_2$ thin film

We started with optimizing the W-film deposition by sputtering, and later, we went through selenization of those films at different optimized growth parameters in the CVD furnace. In this work, we optimized W-film growth using our RF-sputtering set-up with varying parameters listed in the table below.

| RF Power Variation | Ar+H$_2$-gas pressure variation | Process pressure | Optimized RF power | Optimized Gas pressure | Optimized Deposition Rate |
|---|---|---|---|---|---|
| 50-150 W | 0.5-2 kg/cm$^2$ | 1.3*10$^{-3}$ mbar | 100W | 1 kg/cm$^2$ | 5-7 nm/min |

The RF power has been varied in this deposition process for proper plasma sheath formation under varying Ar-gas pressure. It has been observed that deposition properly occurs at power 100W (placed between target and sample boats) under the gas pressure 1 kg/cm$^2$ with a rate ~ 5-7 nm/min confirmed by atomic force microscopy (AFM) and X-ray reflection (XRR) studies. Ultimately, deposition is controlled by the process pressure 1.3x10$^{-3}$ mbar (effective pressure inside the chamber). In this work, growth has been done at atmospheric pressure. Temperature at zone-1 was varied from 300-360 °C for proper evaporation of Se-powder, whereas zone-2 temperature was tuned from 400-550 °C. Before heat initialization of zone-1, we kept zone-2 temperature at 400-450 °C for pre-oxidation of W-film, which serves as a precursor for forming WSe$_2$.

### 2. Reactive Ion Etching of the grown film

To confirm its exact thickness and to get the realization of ultra-thin structures such as bi/tri-layers, we performed reactive ion etching (RIE) using SF$_6$ and O$_2$ gases with flow rate 30 and 10 sccm respectively at RF power ~ 40W upon a patterned window of photoresist (S-1813) created by laser writer. Later weightage of O$_2$ and SF$_6$ were decreased down to 3 and 20 sccm respectively to partially etch the film with reduced RF power ~ 30-15 Watt as depicted by optical microscope image, **Fig.3-(c).**

3. **Device fabrication**

To define the natural doping nature and photo response characteristics of these CVD-grown films, we have measured the capacitance on the fabricated vertical metal-insulator-semiconductor (MOS) structure, followed by comparing the effect arising from the bottom $SiO_2$/p-Si structure and W at the middle. To fabricate such DMOS structure, 10nm $HfO_2$ and Ti:Au (10:60 nm) have been deposited over CVD grown $WSe_2$ using electron beam evaporation at high vacuum pressure ~ $2 \times 10^{-6}$ mbar as schematically presented in **Fig.(5-a).**

## Acknowledgements

The authors thank the Department of Science and Technology (DST) with grant no. CRG/2020/002262 for financial support. The authors express gratitude to the Ministry of Education (MoE), Govt. of India, with grant no. MoE-STARS/STARS-2/2023-0621 for financial support. The authors are grateful to Central Research Facility (CRF), Sophisticated Analytical & Technical Help Institutes (SATHI), Nanoscale Research Facility (NRF), IIT Delhi for infrastructural support. Authors acknowledge Dr. Basudev Nag Chowdhury, IIT Kharagpur and Dr. Atul Kumar Singh, CRF, IIT Delhi for fruitful discussions.

## Authors Contributions

KS and AM performed the sputtering and CVD experiments followed by the materials' characterizations. AM and KS performed structural analysis by DFT and HR-TEM. KS, AM and KB fabricated the devices with corresponding electrical and optical measurements. AM and KS performed laser writer and RIE experiments. BS performed the cross-sectional STEM measurement. AM, KS and KB wrote the manuscript with input from all authors. SD and DM supervised this work.

## Supporting Information

S1: Layer Induced Impact, S2. Interface of W/$WSe_2$ heterostructure, S3. Capacitance 3D-plot for DMOS device, S4. HR-TEM measurements